\documentclass[12pt]{article}
\usepackage[dvips]{epsfig}
\usepackage[dvips]{color}
\usepackage{colordvi}
\newcommand{\Journal}[4]{{\it #1} {\bf #2}, #3 (#4)}
\newcommand{\MISR}[2]{{\cal M}^{(#2)}_{#1}}
\newcommand{\dsISR}[2]{{d\sigma}^{(#2)}_{#1}}
\newcommand{\Lf}{{\rm Lf}}
\newcommand{\Sf}{{\rm Sf}}
\newcommand {\Sp}{{\rm Sp}\;}
\renewcommand{\Re}{{\rm Re}\;}
\newcommand{\refcite}[1]{\cite{#1}}
\begin{document}
\title{NUMERICALLY STABLE CALCULATIONS OF RADIATIVE
CORRECTIONS TO BREMSSTRAHLUNG IN \lowercase{$e^+ e^-$} ANNIHILATION}
\author{S.A. YOST and B.F.L. WARD\\
Department of Physics, Baylor University, Waco, TX 76798, USA}
\maketitle
\kern-4in
\begin{flushright}
 {\bf BU-HEPP-06/09}\\
\end{flushright}
\kern 4in

\abstract{We discuss techniques for obtaining a 
numerically stable evaluation of the fully differential cross section
for the virtual photon correction to single hard photon bremsstrahlung
in two different computational schemes.  We also compare the role of 
finite mass corrections in these schemes.}

\section{Introduction}\label{sec:intro}

Electron-positron annihilation to fermions, $e^+e^- \rightarrow f{\overline f}$,
plays a critical role in extracting precision electroweak physics from 
$e^+e^-$ colliders. Precise calculations are needed for the final 
data analysis of LEP and the anticipated ILC. Thus, the basic process must
be augmented by radiative effects, in particular bremsstrahlung from a
fermion line, $e^+e^- \rightarrow f{\overline f}\gamma$. More photons may
be added as needed, including virtual photons exchanged in internal lines
to the desired order. Single hard bremstrahlung is also of importance in
``radiative return'' experiments, where the radiated photon is used to reduce
the effective energy of the collision, permitting a range of energies to be
studied at a fixed-energy accelerator. This technique is useful at B factories
and high-energy colliders. 

In this note, we will be concerned with the accurate
calculation of one-loop corrections to single real photon emissions, where
there is one extra virtual photon in the process.\cite{jmwy}
We will discuss some aspects of 
calculating these corrections in a manner stable enough to permit 
high-precision Monte Carlo comparisons to related results.\cite{kuhn1,kuhn2}
We will also discuss the role of finite-mass corrections, and compare our 
approach to that of Ref.\ \refcite{kuhn2}, 
examining the role of mass corrections
in the level of agreement found for these results.

\section{Virtual Corrections to Bremsstrahlung}\label{sec:virtual}

The one-loop virtual correction to initial or final-state bremsstrahlung 
was calculated in Ref.\ \refcite{jmwy} using helicity
spinor methods, which provide an efficient representation of massless fermion
scattering, including a ``magic'' choice for the photon polarization vectors
which eliminates many terms from the calculation.\cite{magic1,magic2,magic3} 
The amplitudes were simplified using the symbolic manipulation FORM,\cite{FORM} 
and the scalar one-loop Feynman integrals were evaluated using the FF 
package.\cite{FF} Eventually, these integrals were replaced by the analytic 
expressions in Ref.\ \refcite{jmwy} as shown in the appendix of that paper.  
The amplitudes are then evaluated by the ${\cal KK}$ Monte Carlo 
program,\cite{KKMC} which squares and sums them numerically when creating 
events to obtain the scattering cross section.

The initial state virtual photon contribution to the cross-section may be
expressed as 
\begin{equation}
\label{dsISR11}
\frac{\dsISR{1}{1}}{d\Omega dr_1 dr_2} = \frac{1}{(8\pi)^4}
 \sum_{\lambda_i, \sigma} 2\,{\rm Re} \left[ (\MISR{1}{0})^* \MISR{1}{1}\right] 
\end{equation}
where the tree-level result is 
\begin{equation}
\label{dsISR10}
\frac{\dsISR{1}{0}}{d\Omega dr_1 dr_2} = \frac{1}{(8\pi)^4}
 \sum_{\lambda_i, \sigma} \left| \MISR{1}{0}\right|^2
\end{equation}
with summed-squared matrix element 
\begin{equation}
\label{MISR}
\sum_{\lambda_i, \sigma} \left|\MISR{1}{0}\right|^2
= \frac{16 e^6}{s^2 s' r_1 r_2}
\left\{\left(t_1^2 + u_1^2\right) \left(1 - \frac{2m_e^2 r_1}{s' r_2}\right)
     + \left(t_2^2 + u_2^2\right) \left(1 - \frac{2m_e^2 r_2}{s' r_1}\right)
\right\} ,
\end{equation}
%
with $s = (p_1+p_2)^2$, $s' = (p_3 + p_4)^2$,
$t_1 = (p_1 - p_3)^2$, $t_2 = (p_2 - p_4)^2$,
$u_1 = (p_1 - p_4)^2$, $u_2 = (p_2 - p_3)^2$, $r_i = 2p_i\cdot k/s$, where
$p_1, p_2, p_3, p_4, k$ are the momenta of the $e^-, e^+, f, {\overline f},$
and $\gamma$.
The explicit mass corrections in Eq.\ \ref{MISR} are obtained using
the method of Ref.\ \refcite{berends}, as discussed in Sec.\ 3.
The matrix element for hard photon initial-state
emission with one virtual photon may be expressed as
\begin{equation}
\label{vdef1}
\MISR{1}{1} = \frac\alpha{4\pi} \MISR{1}{0} (f_0 + f_1 I_1 + f_2 I_2),
\end{equation}
where $f_i$ are scalar form factors and $I_i$ are spinor factors defined in
Ref.\ \refcite{jmwy}.  

The expressions for the $f_i$ include
differences between logarithms and dilogarithms with arguments which are 
very similar in collinear limits, and these differences are frequently 
divided by the collinear factors $r_i$, so that the result is finite
in the collinear limits.  Evaluating such expressions in a numerically stable
manner requires expansions where appropriate.  A set of functions
useful for this purpose are the logarithmic and dilogarithmic
difference functions $\Lf_n(x)$ and $\Sf_n(x,y)$ 
introduced in Refs.\ \refcite{paris,compare2} and defined recursively via
\begin{eqnarray}
\label{difdef}
\Lf_0(x) &=& \ln(1+x),  \nonumber\\
\Lf_{n+1}(x) &=& \frac1x \left(\Lf_n(x) - \Lf_n(0)\right),\nonumber\\
\Sf_0(x, y) &=& \Sp(x+y),  \\
\Sf_{n+1}(x, y) &=& \frac1y \left(\Sf_n(x,y) - \Sf_n(x,0)\right),\nonumber
\end{eqnarray}
with $\Sp(x)$ the dilogarithm (Spence function). A set of identities and 
expansions for these functions may be found in the appendix of 
Ref.\ \refcite{compare2}. Thus, for example, the form factor $f_0$ (the 
only term in Eq.\ \ref{vdef1} which survives in the collinear limits), may
be expressed as, without mass corrections, 
\begin{eqnarray}
f_0 &=& 
\left(2\ln\frac{\lambda^2}{m_e^2} + 3 - i\pi\right) 
\left(L - 1 - i\pi\right) 
- L^2 - 1 + \frac{\pi^2}3 \nonumber\\
&+& 
 \frac{r_2(2+r_1)}{(1-r_1)(1-r_2)} \left\{\ln\left(\frac{r_2}{1-v}\right)
+ i\pi\right\} 
+ \frac{r_2}{1-r_2}\nonumber\\
&-& \left\{3v + \frac{2r_2}{1-r_2}\right\} \Lf_1\left(-v\right)
+ \frac{v}{(1-r_2)}\; R_1(r_1,r_2) + r_2 R_1(r_2,r_1)
\end{eqnarray}
with $L = \ln(s/m_e^2)$ the ``big logarithm'' of a leading log expansion,
$v = r_1 + r_2$ the fraction of the beam energy radiated into the hard 
photon, and
\begin{eqnarray}
R_1(x, y) &=& \Lf_1(-x)\left\{\ln\left(\frac{1-x}{y^2}\right) - 2\pi i
\right\}\ \nonumber\\
&+& \frac{2(1-x-y)}{1-x}\ \Sf_1\left(\frac{y}{1-x}, \frac{x(1-x-y)}{1-x}
\right).
\end{eqnarray}
The parameter $\lambda$ is a photon mass cutoff for the infrared divergence. 
The expression for $f_0$ appearing here
and in Ref.\ \refcite{compare2} is analytically identical to the expression in 
Ref.\ \refcite{jmwy}, but is preferable for numerical evaluation, because it can
be evaluated in a stable manner in collinear limits.

\section{Finite Mass Corrections}

It may not be immediately clear that there is any need to consider the finite
mass of the electron in high energy scattering: for the ILC, $m_e^2/s$ is of
order $10^{-12}$.  But in fact, in any process where collinear photon emission 
is possible, the electron mass cannot be neglected, regardless of the 
scattering energy.  This is because integrating terms of the form 
$m_e^2/(p\cdot k)^2$ over photon momentum $k$ always gives contributions of 
order 1.  Obtaining precise results for collinear emission therefore 
requires mass corrections to be added.  Nevertheless, most finite 
mass terms are negligible, so it makes sense to use a massless 
helicity-amplitude technique, but supplement it with a procedure\cite{berends} 
to restore the essential collinear mass terms. 

For ISR, the net result is to add a mass term to the squared amplitude of 
the form 
\begin{equation}
\left|\MISR{1}{m}\right|^2 = 
-\sum_i \frac{e^2 m_e^2}{(p_i\cdot k)^2} 
\Big|{\cal M}_{\rm Born}(p_i - k)\Big|^2,
\end{equation}
where the sum is over the two incoming fermion lines. 
The effect on the form factors is that the spin-averaged value of $f_0$
receives an additional mass term
\begin{eqnarray}\label{fmass}
\langle f_0 \rangle^{(m)} &=& \frac{2m_e^2}{sr_1 r_2}
\frac{(1 - v)(r_1^2 + r_2^2)}{(1-r_1)^2 + (1-r_2)^2}\\
&\times& \left\{\langle f_0 \rangle + \ln\left(\frac{s}{m_e^2}\right)
\left[\ln(1-v) - 1\right]
- \frac32\ln(1-v) + \frac12\ln^2(1-v) + 1\right\}.\nonumber
\end{eqnarray}
This representation of the mass corrections is very compact and well suited
to Monte Carlo integration.  

\section{Comparison of Virtual Photon Corrections to ISR}

Another expression for the virtual photon corrections
to initial state radiation (ISR) has been calculated in 
Refs.\ \refcite{kuhn1,kuhn2}
using a ``leptonic tensor'' and an expansion in factors of $m_e^2/p_i\cdot k$.
This result incorporates the same processes as Ref.\ \refcite{jmwy} and claims
 the same order of exactness, but was obtained by very different means, 
so it provides a valuable cross-check. 

An analytic comparison\cite{compare2,beijing}
showed that the massless parts (without explicit electron mass corrections) of
the results were identical in collinear limits (small $p_i\cdot k$), which
means that the results are identical to NLL order (order $L$) in an expansion
in a leading-log expansion.  In this limit, the functions $f_1$
and $f_2$ in Eq.\ \ref{vdef1} vanish and $f_0$ simplifies greatly. The 
massless spin-averaged NLL limit is 
\begin{eqnarray}\label{NLL}
\Re \langle f_0 \rangle^{\rm NLL} &=&
\left(2\ln\frac{\lambda^2}{m_e^2} + 3\right)(L-1) 
- L^2 - 1 + \frac{4\pi^2}3
+ \frac{r_1(1-r_1)}{1+(1-r_1)^2}\nonumber\\
&+& \frac{r_2(1-r_2)}{1+(1-r_2)^2} + 2\ln r_1 \ln(1-r_2) 
+ 2\ln r_2 \ln(1-r_1) \\
&-& \ln^2(1-r_1) 
- \ln^2(1-r_2) + 3\ln(1-r_1) + 3\ln(1-r_2),\nonumber
\end{eqnarray}
and mass corrections can be added via the collinear limits of Eq.\ \ref{fmass}.

The mass corrections are more difficult to compare analytically, both
because the mass corrections of Ref.\ \refcite{kuhn2} are much longer, and
because those expressions include 
terms proportional to $m_e^4(p_i\cdot k)^{-3}$ 
which are absent in Ref.\ \refcite{jmwy}.  Rewriting the mass corrections
using the functions $\Lf_n$ and $\Sf_n$ shows that all such factors actually 
cancel, and the two expressions for the mass corrections agree to NLL 
order,\cite{compare2} in the sense that all terms producing at least one 
factor of $L$ upon integration agree.  In the case of mass corrections, this 
statement is not as strong as saying that the collinear limits agree. 
In fact, we have shown that the 
soft collinear limits agree analytically, but the $L^0$ hard collinear limits 
are not identical.  

\begin{figure}[ht]
\centerline{\includegraphics[width=2.5in]{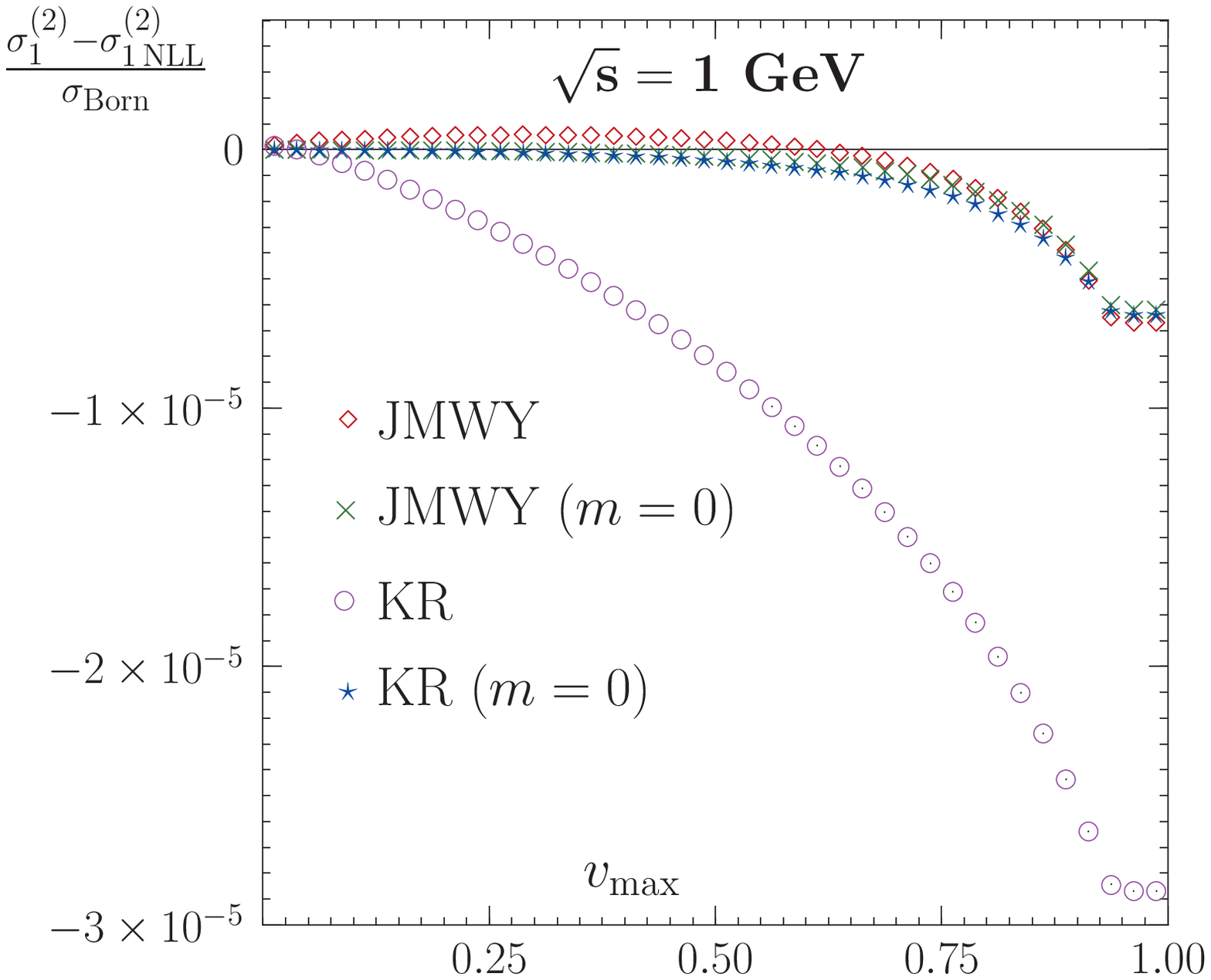}$\qquad$
	    \includegraphics[width=2.5in]{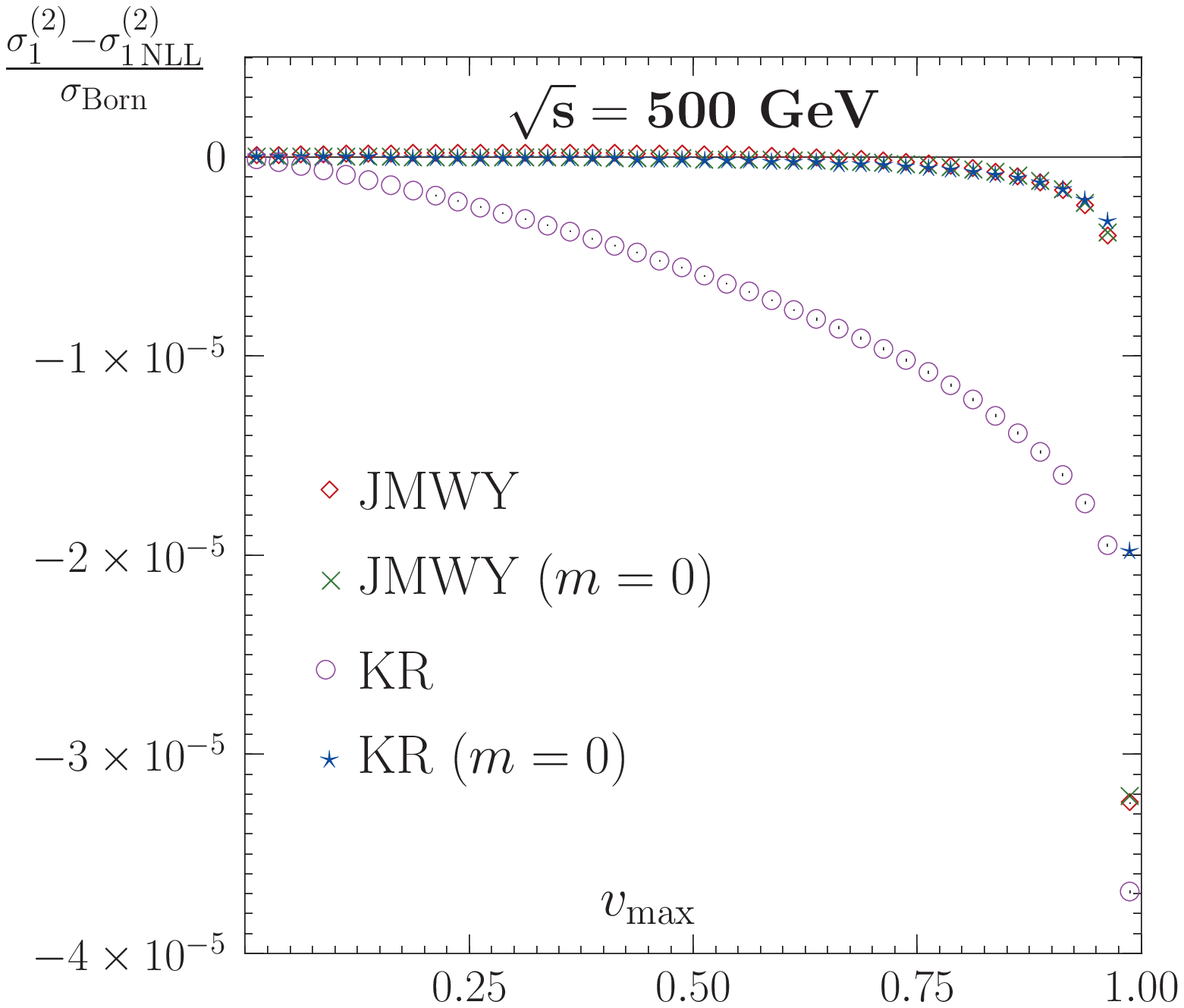}}
\caption{\footnotesize
Differences between the exact and massless NLL results for the 
virtual correction to single photon ISR, both with and
without mass corrections, in a ${\cal KK}$MC run of $10^8$ events. 
Results are shown for two different energy scales, $\sqrt{s} = $
1 GeV and 500 GeV.  JMWY is the expression of 
Ref.\ \protect\refcite{jmwy} and KR is the expression of 
Ref.\ \protect\refcite{kuhn2}, and $v_{\max}$ is the 
limit on the radiated photon energy, relative to the beam energy.}
\end{figure}

Detailed numerical comparisons\cite{compare2,epiphany,compare1}
 have been made using the ${\cal KK}$MC, 
both at 1 GeV and 500 GeV, corresponding to typical energies for 
B factories and the ILC, respectively.  Since the massless NLL limits 
are known to agree, we have subtracted the NLL limit (Eq.\ \ref{NLL}) in each
case, and plotted the remaining contribution, both with and 
without explicit mass terms.  The results of MC runs for $10^{8}$ events are 
shown in Fig.\ 1, normalized to the Born cross section.  

The plots show that the massless results agree to within $10^{-6}$ for all but
the last bin of the 500 GeV run.  The main difference is evident in the mass 
corrections. The maximum difference between the integrated cross sections
is found to be $2.2\times 10^{-5}$ at 1 GeV and $1.6\times 10^{-5}$ at 500 GeV,
in units of the Born cross section.\footnote{For radiative return, it is more relevant to compare to the hard ISR cross-section $\sigma_1$, which is 0.113 $\sigma_{\rm Born}$ at 1 GeV and 0.980 $\sigma_{\rm Born}$ at 500 GeV.}  
While this is an excellent level of 
agreement, it is clear that there is some disagreement in the mass corrections,
which are responsible for almost the entire difference.  It would be desirable 
to understand the origin of this difference more completely.

\section*{Acknowledgments}
This work was supported in part by U.S. Department of Energy grant
DE-FG02-05ER41399.

\end{document}